\documentclass[prb,reprint,amsmath,amssymb,aps,twocolumn,superscriptaddress]{revtex4-2}
\usepackage{hyperref}
\usepackage{rotating}
\usepackage{graphicx}
\usepackage{latexsym}
\usepackage{amssymb}
\usepackage{amsfonts}
\usepackage{soul}
\usepackage{amsmath}
\usepackage{color}
\usepackage{lipsum}
\usepackage{bbm}

\DeclareMathAlphabet\mathbfcal{OMS}{cmsy}{b}{n}

\newcommand{\vecA}{\mathbfcal{A}}
\newcommand{\vecB}{\mathbfcal{B}}

\newcommand{\vecD}{\mathbfcal{D}}
\newcommand{\cbb}{\mathbbmss{c}}
\newcommand{\fbb}{\mathbbmss{f}}

\newcommand{\gbb}{\mathbbmss{g}}

\newcommand{\grm}{\mathrm{g}}
\newcommand{\frm}{\mathrm{f}}

\begin{document}

\title{Intertype superconductivity evoked by the interplay of disorder and multiple bands}

\author{P. M. Marychev}
\email{pmarychev@hse.ru}
\affiliation{HSE University, 101000 Moscow, Russia}

\author{A. A. Shanenko}
\affiliation{HSE University, 101000 Moscow, Russia}

\author{A. Vagov}
\affiliation{HSE University, 101000 Moscow, Russia}

\date{\today}
\begin{abstract}
Nonmagnetic impurity scattering is known to shift up the Ginzburg-Landau parameter $\kappa$ of a superconductor. In this case, when the system is initially in type I, it can change its magnetic response, crossing the intertype domain with $\kappa \sim 1$ between the two standard superconductivity types and arriving at type II. In the present work we demonstrate that the impact of disorder can be much more profound in the presence of the multiband structure of the charge carrier states. In particular, when the band diffusivities differ from each other, the intertype domain tends to expand significantly, including points with $\kappa \gg 1$ that belong to deep type-II in conventional single-band superconductors. Our finding sheds light on the nontrivial disorder effect and significantly complements earlier results on the enlargement of the intertype domain in clean multiband superconductors. 
\end{abstract}

\maketitle

It is well-known that a nonmagnetic disorder can influence the superconductive magnetic properties by altering the characteristic lengths of a superconductor~\cite{Ketterson}. In particular, the Ginzburg-Landau (GL) coherence length $\xi$ decreases when the electron mean-free path is reduced. At the same time the London magnetic penetration depth $\lambda$ increases. As a result, the ratio $\kappa=\lambda/\xi$, referred to as the GL parameter, increases as well. In this case the system, when being initially in type I, crosses the intertype (IT) domain between the two standard superconductivity types in the $\kappa$-$T$ plane ($T$ is the temperature) and exhibits the type-II magnetic response at a sufficient disorder. This feature was used to study the IT superconductivity and its boundaries when the magnetic properties of, e.g., Ta and Nb were modified by changing the amount of dissolved nitrogen~\cite{Auer1973}. The IT superconductivity is of special interest since it is characterized by unconventional magnetic properties and flux-condensate distributions which differ qualitatively from those of the two standard superconductivity types. A number of studies~\cite{Krageloh1969, Essmann1971, Kumpf1971, Jacobs1971A, Ovchinnikov1999, Luk'yanchuk2001, Laver2009, Brandt2011, Pautrat2014, Ge2014, Reimann2015, Vagov2016, Ge2017, Reimann2017, Wolf2017A, Backs2019, Saraiva2019, Vagov2020, Ooi-2021, Brems2022} demonstrated that for conventional materials, the IT physics manifests itself at $\kappa \sim 1$. 

In the present work we report a striking example when disorder does not only shift the system across the IT regime. Here the interplay of the diffusive motion of charge carriers with the multiband structure of the carrier states leads to qualitative changes in the magnetic-response phase diagram. When the band diffusivities differ significantly from each other (as e.g. in films of MgB$_2$~\cite{Curran2015}), the IT domain shows a giant expansion. As a result, it can include large values of the GL parameter ($\kappa \gg 1$) that belong to deep type II in conventional single-band superconductors. Our finding significantly complements earlier results on the enlargement of the IT domain in clean multiband superconductors~\cite{Vagov2016, Wolf2017B, Cavalcanti2020}. 

As the prototype of a multiband superconductor we choose the two-band system with the $s$-wave pairing in both bands and Josephson-like interband pair transfer. To describe the corresponding IT domain in the dirty limit, we employ the two-band Usadel equations~\cite{Gurevich2003}. To avoid unnecessary complications of an anisotropic case, the system is assumed to be isotropic. For simplicity we neglect the interband impurity scattering since our preliminary results demonstrates that it can produce quantitative corrections but does not change the qualitative picture. Investigations of such corrections will be published elsewhere. Then, the equations for the band-dependent gap functions $\Delta_{\nu}=\Delta_{\nu}({\bf r})$ read ($\nu=1,2$)
\begin{align}
\hbar\omega f_{\nu}-\frac{\hbar {\cal D}_{\nu}}{2}\big(g_{\nu}\boldsymbol{D}^2f_{\nu}-f_{\nu}\boldsymbol{\nabla}^2g_{\nu}\big)=\Delta_{\nu} g_{\nu},
\label{usadel}
\end{align}
where $g_{\nu}=g_{\nu}({\bf r},\omega)$ and $f_{\nu}=f_{\nu}({\bf r},\omega)$ are the normal and anomalous quasiclassic (frequency-dependent) Green functions related to one another by the normalization condition $g^2_{\nu} +|f_{\nu}|^2=1$, $\omega$ stands for the fermionic Matsubara frequencies, ${\cal D}_{\nu}$ is the diffusion coefficient associated with band $\nu$, and ${\bf D}=\boldsymbol{\nabla}-(i2e/\hbar \cbb){\bf A}$ is the gauge-invariant derivative. The Usadel equations (\ref{usadel}) are solved together with the self-consistency relation
\begin{align}
\Delta_{\nu}= 2\pi T \sum_{\nu'=1,2}  \grm_{\nu\nu'} N_{\nu'}\sum_{\omega>0} f_{\nu'},
\label{self}
\end{align}
where $\grm_{\nu\nu'}$ is the element of the symmetric coupling matrix $\check{\grm}$ and $N_{\nu}$ is the band density of states (DOS). 

The free energy density for the system of interest is given by 
\begin{align}
\fbb =\frac{{\bf B}^2}{8\pi}+\langle \vec\Delta^{\dagger},\check{\grm}^{-1}\vec{\Delta}\rangle +\sum\limits_{\nu=1,2}  \fbb_{\nu}, \label{freeen}
\end{align}
where ${\bf B}=\boldsymbol{\nabla}\times{\bf A}$ is the magnetic field, $\vec{\Delta}=(\Delta_1,\Delta_2)^T$ with $\langle .,. \rangle$ the scalar product in the band vector space, $\check{\grm}^{-1}$ is the inverse of the coupling matrix and 
\begin{align}
\fbb_{\nu}= &2\pi T N_{\nu}\sum_{\omega>0}\Big\{ 2\hbar\omega (1-g_{\nu})-2\mathrm{Re}(f_{\nu}^*\Delta_{\nu})\notag\\
&+\frac{\hbar\cal{D}_{\nu}}{2}\big[|{\bf D} f_{\nu}|^2+(\boldsymbol{\nabla} g_{\nu})^2\big]\Big\}.
\label{fnuA}
\end{align}
The stationary point (minimum) of the free energy gives the equilibrium spatial distributions of $\Delta_1({\bf r})$, $\Delta_2({\bf r})$ and ${\bf B}({\bf r})$~[and ${\bf A}({\bf r})$]. 

To calculate the boundaries of the IT domain, we employ the perturbation expansion of the two-band Usadel formalism in the small deviation from the superconducting critical temperature $\tau = 1- T/T_c$. It was shown previously for clean two-band superconductors~\cite{Vagov2016,Wolf2017A,Vagov2020} that many important details regarding the intertype superconductivity can be obtained already from the leading correction to the GL theory in $\tau$. The derivation of this correction in the present case is similar to that for clean two-band superconductors~\cite{Vagov2016, Wolf2017A, Saraiva2019, Vagov2020, Cavalcanti2020}. (For general details of the $\tau$-expansion in single- and multiband superconductors, see the papers \cite{Jacobs1971A, Jacobs1971B, Vagov2012A} and \cite{Shanenko2011,Vagov2012B}, respectively.) First, one represents the Green functions and the free energy density as series in powers of the gap functions and their spatial derivatives. The series are truncated so that to keep only the terms that contribute to the leading correction to the GL theory in $\tau$. Second, based on the obtained expressions, one derives the $\tau$-expansion of the formalism up to the leading correction to the GL theory.  

Now, we employ the Usadel equations and invoke the expansion in powers of the gap functions and their spatial gradients (for more detail, see the Supplementary material). For the free energy density one gets 
\begin{align}
\fbb_{\nu}= &\Big[-N_{\nu}A +a_{\nu}\left(\tau+\frac{\tau^2}{2}\right)\Big] |\Delta_{\nu}|^2+\frac{b_{\nu}}{2}(1+2\tau)|\Delta_{\nu}|^4\notag\\
&-\frac{c_{\nu}|\Delta_{\nu}|^6}{3}+\mathcal{K}_{\nu}(1+\tau)|{\bf D}\Delta_{\nu}|^2 - \mathcal{Q}_{\nu}|{\bf D}^2\Delta_{\nu}|^2\notag\\
&-\frac{\mathcal{L_{\nu}}}{2}\Big\{6|\Delta_{\nu}|^2|{\bf D}\Delta_{\nu}|^2+\big[\Delta_{\nu}^2({\bf D}^*\Delta_{\nu}^*)^2+ {\rm c.c.}\big]\Big\},
\label{fnuB}
\end{align}
where 
\begin{align}
\nonumber
&a=- N_{\nu}, A=\ln\frac{2e^{\gamma}\hbar\omega_D}{\pi T_c}, b_{\nu}=N_{\nu}\frac{7\zeta (3)}{8\pi^2T_c^2},\notag\\ 
&\mathcal{K}_{\nu}=N_{\nu}\frac{\pi\hbar\mathcal{D}_{\nu}}{8T_c}, \mathcal{Q}_{\nu}=\frac{(\hbar\mathcal{D}_{\nu})^2}{2} b_{\nu},\notag\\
&\mathcal{L}_{\nu}=N_{\nu}\frac{\pi \hbar \mathcal{D}_{\nu}}{192T_c^3}, c_{\nu}=N_{\nu}\frac{93\zeta(5)}{128\pi^4T_c^4},
\label{coeffs}
\end{align}
with $T_c$ the critical temperature, $\omega_D$ the Debye frequency, $\zeta(\ldots)$ the Riemann zeta function, and $\gamma=0.577$. As is mentioned above, the series in Eq.~(\ref{fnuB}) is truncated here so that to include only the terms that contribute to the leading correction to the GL theory in $\tau$. 

It is instructive to compare the free energy density given by Eqs.~(\ref{freeen}) and (\ref{fnuB}) for the two-band system in the dirty limit with the corresponding expansion of the free energy density in the clean limit~\cite{Shanenko2011, Vagov2012B, Vagov2016, Cavalcanti2020}. First, the coefficients $\mathcal{K}_{\nu},\mathcal{Q}_{\nu}$, and $\mathcal{L}_{\nu}$ are now given by the different expressions [see Eq.~(\ref{coeffs})] including the band-dependent diffusivities $\mathcal{D}_{\nu}$. Second, the set of the three terms with the coefficient $\mathcal{Q}_{\nu}$, calculated for the clean limit, is now reduced to the only term in Eq.~(\ref{coeffs}): there are no contributions proportional to $\boldsymbol{\nabla}\times{\bf B}$ and ${\bf B}^2$, c.f. Eq.~(\ref{fnuB}) with Eq.~(20) in \cite{Vagov2012B}. Finally, the first term in the figure braces in the last line of Eq.~(\ref{fnuB}) has now a numerical factor $6$ instead of $8$ in the clean limit. However, in general, the structure of the free energy density given by Eqs.~(\ref{freeen}) and (\ref{fnuB}) is similar to that of the clean system. Thus, to obtain the $\tau$-expansion of the present microscopic formalism, we can employ a similar calculation procedure.  

Based on the previous results for clean two-band superconductors~\cite{Shanenko2011, Vagov2012B, Vagov2016}, we introduce the $\tau$-expansion for the gap functions and fields in the form
\begin{align}
    &\vec{\Delta} = \tau^{1/2} \vec{\Psi}+ \tau^{3/2}\vec{\psi}+\dots, \;{\bf B} = \tau \vecB + \tau^2  \boldsymbol{\mathfrak{b}} + \dots,\notag\\
    &{\bf A} = \tau^{1/2} \vecA + \tau^{3/2} \boldsymbol{\mathfrak{a}} + \dots, 
\label{tau}
\end{align} 
where $\vec{\Psi}=(\Psi_1,\Psi_2)^T$ and $\vecB$~($\vecA$) correspond to the GL theory while its leading correction is governed by $\vec{\psi}=(\psi_1,\psi_2)^T$ and $\mathfrak{b}$~($\mathfrak{a}$). In addition, the magnetic penetration depth $\lambda$ and the GL coherence length $\xi$ are divergent as $\lambda, \xi \propto \tau^{-1/2}$. To extract this dependence from the spatial gradients, we introduce the spatial scaling ${\bf r} \to \tau^{1/2}{\bf r}$ and obtain the corresponding scaling factor for the spatial derivatives as $\boldsymbol{\nabla} \to \tau^{-1/2} {\boldsymbol{\nabla}}$. Then, based on Eqs.~(\ref{freeen})-(\ref{tau}) [see also the Supplementary material], one gets the stationary equations for $\vec{\Psi}$ and $\vec{\psi}$ as
\begin{align}
\check{L}\vec{\Psi} = 0,\quad \check{L}=\check{\grm}^{-1}
-\left(\begin{array}{cc} N_1 A & 0\\
0 & N_2 A
\end{array}
\right) 
\label{gap0}
\end{align}
and 
\begin{align}
\check{L}\vec{\psi} +\vec{W} = 0, 
\label{gap1}
\end{align}
where $\vec{W}=(W_1,W_2)^T$ and $W_{\nu}= a_{\nu}\Psi_{\nu}+\frac{b_{\nu}}{2}\Psi_{\nu}|\Psi_{\nu}|^2 +{\cal K}_{\nu} \vecD^2\Psi_{\nu}$, with $\vecD=\boldsymbol{\nabla}-(i2e/\hbar \cbb)\vecA$. 

Equation (\ref{gap0}) has a nontrivial solution when the determinant of the matrix $\check{L}$ is zero and we obtain
\begin{equation}
(\grm_{22}- \mathcal{G} N_1 A )(\grm_{11}- \mathcal{G} N_2 A)-\grm^2_{12}=0,\; \vec{\Psi} =\Psi({\bf r})\vec{\xi},
\label{Tc}
\end{equation}
where $\mathcal{G}=\grm_{11}\grm_{22}-\grm^2_{12}$, $\Psi$ is the Landau order parameter that controls the two-band system in the GL approximation, and $\vec{\xi}$ is the eigenvector of $\check{L}$ corresponding to its zero eigenvalue. The normalization of $\vec{\xi}$ is not important here (the observables are not sensitive to it) and so, there are various options to choose $\vec{\xi}$. Here we follow the variant used in \cite{Vagov2016} and given by  
\begin{align}
\vec{\xi}=
\left(\begin{array}{c}
S^{-1/2}\\
S^{1/2}
\end{array}\right), \; S=\frac{\grm_{22} - \mathcal{G} N_1 A}{\grm_{12}},
\label{xi}
\end{align}
where $S$ controls the relative weights of the bands, changing from $0$ (only band 2) to $\infty$ (only band 1). 

Introducing the vector 
\begin{align}
\vec{\eta}=
\left(\begin{array}{c}
S^{-1/2}\\
-S^{1/2}
\end{array}\right)
\label{eta}
\end{align}
so that $\vec{\xi}$ and $\vec{\eta}$ are linearly independent, one can represent $\vec{\psi}$ as their linear combination given by
\begin{align}
\vec{\psi} = \psi_{\xi}({\bf r}) \vec{\xi} + \psi_{\eta}({\bf r}) \vec{\eta},
\label{psi}
\end{align}
where $\psi_{\xi}$ and $\psi_{\eta}$ control the spatial distributions of the gap functions in the leading correction to the GL theory. Projecting Eq.~(\ref{gap1}) onto $\vec{\xi}$ and utilizing Eq.~(\ref{psi}), one gets the GL equation for the Landau order parameter as
\begin{align}
a\Psi + b \Psi|\Psi|^2 -{\cal K}\vecD^2\Psi=0, 
\label{GL1}
\end{align}
where the coefficients $a=\sum_{\nu}|\xi_{\nu}|^2 a_{\nu},\,{\cal K}=\sum_{\nu} |\xi_{\nu}|^2 {\cal K}_{\nu}$, and $b=\sum_{\nu}|\xi_{\nu}|^4 b_{\nu}$ are averages over the contributing bands, with $\xi_1=S^{-1/2}$ and $\xi_2=S^{1/2}$. 

Projecting Eq.~(\ref{gap1}) onto $\vec{\eta}$ and keeping in mind Eq.~(\ref{psi}), we express $\psi_{\eta}$ in terms of $\Psi$ as
\begin{align}
\psi_{\eta} = -\frac{\mathcal{G}}{4g_{12}} \big(\alpha \Psi + \beta \Psi|\Psi|^2 - \Gamma \vecD^2 \Psi\big),
\label{psieta1}
\end{align}
with the coefficients $\alpha=\sum_{\nu} \eta^*_{\nu}\xi_{\nu} a_{\nu},\, \Gamma =\sum_{\nu} \eta^*_{\nu}\xi_{\nu} \mathcal{K}_{\nu}$, and $\beta=\sum_{\nu} \eta^*_{\nu}\xi_{\nu}|\xi_{\nu}|^2 b_{\nu}$~[here $\eta_1=S^{-1/2}$ and $\eta_2=-S^{1/2}$]. Using Eq.~(\ref{GL1}), one can rearrange Eq.~(\ref{psieta1}) as
\begin{align}
\psi_{\eta} = -\frac{\mathcal{G}}{4g_{12}} \big(a\bar{\alpha} \Psi + b\bar{\beta} \Psi|\Psi|^2\big),
\label{psieta2}
\end{align}
with $\bar{\alpha}=\frac{\alpha}{a}-\frac{\Gamma}{\cal K}$ and $\bar{\beta}=\frac{\beta}{b}-
\frac{\Gamma}{\cal K}$. Notice that $\psi_{\eta}({\bf r})$ is responsible for the difference between the spatial profiles of $\Delta_1({\bf r})$ and $\Delta_2({\bf r})$, i.e. it determines the deviation of the band-dependent coherence lengths $\xi_1$ and $\xi_2$ from the GL coherence length $\xi$, see the discussion in \cite{Vagov2016, Cavalcanti2020}. 

The leading correction to the GL contribution in the free energy density does not involve the terms depending on $\psi_{\xi}$. Thus, to calculate the free energy within the extended GL formalism, involving the GL contribution and its leading correction in $\tau$, one needs to know only the solution to the GL formalism (as $\psi_{\eta}$ depends on $\Psi$ and $\vecA$), see the details in the previous papers for clean two-band systems~\cite{Vagov2016, Cavalcanti2020}. The Landau order parameter obeys the first GL equation given by Eq.~(\ref{GL1}). The second GL equation (the current equation) reads
\begin{align}
\boldsymbol{\nabla} \times \vecB = \frac{4\pi}{\cbb} \boldsymbol{\mathfrak{j}}, \; 
\boldsymbol{\mathfrak{j}} = \frac{4e {\cal K}}{\hbar}{\rm Im} [\Psi^*\vecD \Psi].
\label{GL2}
\end{align}
Using solutions for Eqs.~(\ref{GL1}) and (\ref{GL2}) and employing Eq.~(\ref{psieta2}), one gets the stationary free energy density necessary to investigate the IT physics in dirty two-band superconductors.

Now we turn to the problem of switching between superconductivity types I and II. It is well-known that type II differs from type I by the possibility to develop the mixed state where a magnetic field penetrates the interior of a superconductor so that the superconducting condensate is specified by a nonuniform spatial distribution. To calculate the boundary between types I and II, one needs to compare the Gibbs free energy of the Meissner state at the thermodynamic critical field $H_c$ with that of a specific spatial configuration of the superconducting condensate~\cite{Vagov2016, Cavalcanti2020}. For example, one can choose the single-vortex configuration and calculate the corresponding Gibbs free energy difference between the nonuniform and Meissner states. When this difference is positive, the system is in type I. When it is negative, we arrive at type II. There are several ways to calculate the set of the parameters corresponding to the boundary between types I and II. Within the GL theory all these ways yield the same result: the boundary between types I and II is specified by the relation $\kappa= \kappa_0=1/\sqrt{2}$.

This is not the case beyond the GL theory: here the above ways of calculating the boundary between types I and II result in different lines $\kappa^*(T)$ in the $\kappa$-$T$ plane. All these lines intersect at the point $(\kappa_0,T_c)$, which is called the Bogomolnyi point (B-point). When the system approaches the B-point, it is governed by the self-dual GL theory given by the two Bogomolnyi (self-duality) equations. The fundamental feature of the B-point is that the corresponding equilibrium state is degenerate, hiding an infinite number of various exotic vortex configurations being degenerate solutions of the Bogomolnyi equations~\cite{Vagov2020}. Below $T_c$ the degeneracy is lifted and successive self-dual configurations shape the internal structure of the IT domain and determine its unconventional superconductive magnetic properties~\cite{Vagov2016, Wolf2017A, Cavalcanti2020, Vagov2020}. 

The difference between the Gibbs free energies of a nonuniform condensate configuration and the Meissner state writes as 
\begin{align}
G = \int \gbb \, d^3{\bf r} , \quad \gbb = \fbb +\frac{H^2_c}{8\pi}- \frac{H_c B}{4\pi},
\label{Gibbs_en}
\end{align}
with the applied and internal fields ${\bf H} = (0,0,H_c)$ and ${\bf B} = (0,0,B)$. Here the thermodynamic critical field is given by
\begin{align}
\frac{H_c}{\tau\mathcal{H}_c}=1 - \tau \left(\frac{1}{2}+\frac{ac}{3b^2}+\frac{\mathcal{G}a}{4g_{12}}(\bar{\alpha}-\bar{\beta})^2\right) +\ldots,
\label{Hc}
\end{align}
with the GL thermodynamic critical field $\mathcal{H}_c=\sqrt{4\pi a^2/b}$ and $c=\sum_{\nu}|\xi_{\nu}|^6 c_{\nu}$. Notice that $\mathcal{H}_c$ should be multiplied by $\tau$ to get back to the standard definition of the GL thermodynamic critical field.  

\begin{figure*}[!ht]
\centering
\includegraphics[width=0.9\linewidth]{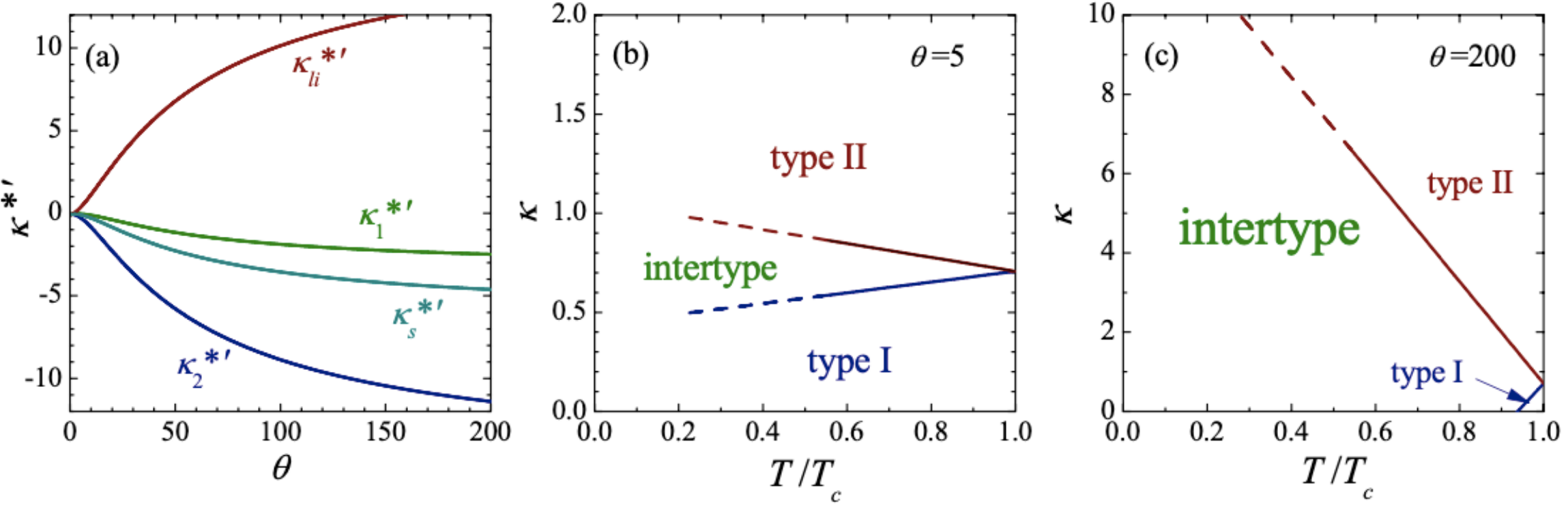}
\caption{The IT domain in the $\kappa$-$T$ phase diagram. Panel (a) demonstrates the $\tau$-derivatives of the GL critical parameters (for their definitions, see the text) versus the ratio $\theta= \mathcal{D}_2/\mathcal{D}_1$. Panels (b) and (c) show the IT domain in the $\kappa$-$T$ plane for $\theta=5$ and $200$; the upper boundary is given by $\kappa^*_{li}(T)$ whereas the lower boundary is $\kappa^*_{2}(T)$. We remark that for single-band superconductors, the experimental results for the boundaries of the IT domain are in good agreement with the calculations of the extended GL theory down to $T \sim 0.5T_c$~\cite{Vagov2016}, this is why our results in panels (b) and (c) are given by the dashed lines below $T=0.5T_c$.} 
\label{fig1}
\end{figure*}

Using the $\tau$-expansion approach, we represent ${\cal G}$ as a series in $\tau$, and keep only the leading correction to the GL contribution (see the Supplemental material). In addition, we employ the expansion in the small deviation $\delta\kappa = \kappa - \kappa_0$, as our study is focused on the IT domain near $\kappa_0$. The relevant details are similar to those in the calculations for clean two-band superconductors and can be found in \cite{Vagov2016, Cavalcanti2020}. Then, the Gibbs free energy difference is obtained as  
\begin{align}
\frac{G}{\tau^2}= & - \sqrt{2}\,  {\cal I} \, \delta \kappa + \tau \, \Big\{ \big[\bar{\cal Q}-\bar{c} + \bar{\mathcal{G}}\bar{\beta}(2\bar{\alpha}-\bar{\beta})\big]{\cal I} \notag \\
& + \Big[\frac{3}{2}{\cal L}-\bar{c}- \bar{\cal Q} - \bar{\mathcal{G}} \bar{\beta}^2\Big] {\cal J}\Big\}+\ldots.
\label{Gibbs_exp}
\end{align}
where $G$ is given in units of $\mathcal{H}_c^2 \lambda^2 L/2\pi$, with $L$ the system size in the $z$ direction, the dimensionless coefficients are defined as
\begin{align}
\bar{c}=\frac{ca}{3b^2},\;\bar{\mathcal{Q}}=\frac{\mathcal{Q}a}{\mathcal{K}^2}, \; \bar{\cal L}=\frac{\mathcal{L} a}{\mathcal{K}b},\; \bar{\mathcal{G}}=\frac{\mathcal{G}a}{4g_{12}},
\label{coeffdimless1}
\end{align}
with $\mathcal{Q}=\sum_{\nu}|\xi_{\nu}|^2 \mathcal{Q}_{\nu},\, \mathcal{L}=\sum_{\nu}|\xi_{\nu}|^4 \mathcal{L}_{\nu}$. The integrals ${\cal I}$ and ${\cal J}$ are given by 
\begin{align}
{\cal I} =\!\! \int\! |\Psi|^2 \big(1 - |\Psi|^2\big)d^2{\bf r},\, {\cal J} =\!\! \int\! |\Psi|^4 \big(1 - |\Psi|^2\big)d^2{\bf r}, 
\label{IJ}
\end{align}
where $\Psi$ is a solution of the GL equations for a particular condensate configuration at $\kappa = \kappa_0$, it is given in units of $\Psi_0=\sqrt{-a/b}$. Using Eq.~(\ref{Gibbs_exp}), we find the corresponding critical GL parameter from $G=0$ as
\begin{align}
\kappa^*= &\kappa_0\bigg\{1 + \tau \, \bigg[\bar{\cal Q}-\bar{c} + \bar{\mathcal{G}}\bar{\beta}(2\bar{\alpha}-\bar{\beta})\notag \\
& + \bigg(\frac{3}{2}\bar{\cal L}-\bar{c}- \bar{\cal Q} - \bar{\mathcal{G}} \bar{\beta}^2\bigg) \frac{\cal J}{\cal I}\bigg]+\ldots\bigg\}.
\label{kappa_exp}
\end{align}
Utilizing a particular condensate-field configuration, we can now find the corresponding critical GL parameter, taking account of the leading correction to the GL theory. 

Notice that the dimensionless GL formalism involves only one parameter, i.e. the GL parameter $\kappa$. It means that for any particular mixed-state configuration taken at $\kappa=\kappa_0$ the spatial distribution of $\Psi$ is the same in both the clean and dirty limits. Then, in our subsequent analysis we can employ the values of ${\cal I}$ and ${\cal J}$ found previously for the clean two-band case. 

One of the possibilities to calculate the boundary between types I and II is to consider the appearance/disappearance of a nonuniform (mixed) superconducting state for the fields above $H_c$. As such states exist below the upper critical field $H_{c2}$, it means that we need to check the condition $H_c=H_{c2}$. In this case $\Psi \to 0$ and so, to get the corresponding critical GL parameter $\kappa^*_2$, one needs to utilize ${\cal J}  \ll {\cal I}$~\cite{Vagov2016} in Eq.~(\ref{kappa_exp}). 

We can also choose the single-vortex solution as the reference spatial configuration and check when it is favourable versus the Meissner state. This is equivalent to the condition $H_c=H_{c1}$, where $H_{c1}$ is the lower critical field~\cite{Vagov2016}. Inserting the corresponding ratio ${\cal J}/{\cal I}= 0.735$~\cite{Vagov2016} in Eq.~(\ref{kappa_exp}), we find the critical parameter $\kappa^*_1$.  

When using the condition of the zero surface energy of a flat domain wall between the superconductive and normal states, one finds ${\cal J}/{\cal I} = 0.559$~\cite{Vagov2016}. This ratio is then plugged in Eq.~(\ref{kappa_exp}), which yield $\kappa^*_s$.  

Finally, there exists also the useful condition of changing the sign of the long-range interaction between vortices. This condition results in ${\cal J}/{\cal I} = 2$~\cite{Vagov2016}, and adopting this ratio in Eq.~(\ref{kappa_exp}), we obtain $\kappa^*_{li}$.  

As these critical GL parameters differ from one another, they yield different boundaries between types I and II beyond the GL theory. This difference shapes the internal structure of the IT domain in the $\kappa$-$T$ plane. To find these boundaries, one needs to explicitly calculate the dimensionless coefficients defined in Eq.~(\ref{coeffdimless1}). These coefficients depend on the three parameters: $S$ given by Eq.~(\ref{xi}) and the ratios $\theta= \mathcal{D}_2/\mathcal{D}_1$ and $\chi= N_2/N_1$. Equation~(\ref{xi}) yields 
\begin{align}
S=\frac{1}{2\lambda_{12}}\!\left[\lambda_{22}-\frac{\lambda_{11}}{\chi} + 
\sqrt{\Big(\lambda_{22}-\frac{\lambda_{11}}{\chi}\Big)^2\!\!+
4\frac{\lambda^2_{12}}{\chi}}\right],  
\label{S}
\end{align}
where $\lambda_{\nu \nu^\prime } = g_{\nu \nu^\prime }(N_1 +  N_2)$. Hence, to get the boundaries of the IT domain in the $\kappa$-$T$ plane, we need to specify the dimensionless couplings $\lambda_{ij}$ together with $\chi$ and $\theta$. Below, for the sake of illustration, we use the set $\lambda_{11}=1.91, \,\lambda_{22}=0.477\, \lambda_{12}=0.204$, and $\chi=1.37$. These values are extracted from the data used for MgB$_2$~\cite{Golubov2002}. The ratio of the band diffusivities is treated as a free parameter here. We remark that this ratio can be very large, up to $\sim 200$, as in dirty films of MgB$_2$~\cite{Curran2015}. It is important to note that the choice of the dimensionless couplings and the DOS ratio is not decisive for our conclusions, similar results are obtained for other variants.  

Our results for $\kappa^*_2,\kappa^*_1,\kappa^*_s$ and $\kappa^*_{li}$ are shown in Figs.~\ref{fig1}(a-c). In Fig.~\ref{fig1}(a) one can see the $\tau$-derivatives of the critical GL parameters as functions of $\theta$. In Figs.~\ref{fig1}(b) and (c) the upper and lower boundaries of the IT domain ($\kappa^*_{li}$ and $\kappa^*_{2}$) are shown in the $\kappa$-$T$ plane for $\theta=5$ and $200$, respectively. The main result of our present investigation is that the IT domain systematically expands with increasing the ratio of the band diffusivities $\theta$. Being nearly negligible at $\theta \sim 1$~[see Fig.~\ref{fig1}(a)], it occupies a significant part of the phase diagram for large values of $\theta$. For example, from Fig.~\ref{fig1}(c) one can see that our diffusive two-band system with $\kappa=$ $7$--$8$ belongs to the IT domain at $T=0.5T_c$ while such $\kappa$-values are commonly thought to be in type II.

For $\theta \lesssim 1$ the IT domain is nearly negligible with the width of about $\Delta\kappa \sim 0.01$~(invisible in the figure) and its upper boundary is given by $\kappa^*_2$. This is similar to the IT domain in a diffusive single-band system, where $H_{c2} < H_c$, and the first order transition is expected at the upper critical field~\cite{Ovchinnikov1999, Luk'yanchuk2001}. However, for $\theta > 3$ the situation changes qualitatively so that the upper IT boundary corresponds to the sign change of the long-range interaction between vortices (controlled by $\kappa^*_{li}$), similarly to the IT domain in clean single-band and two-band superconductors. Based on the previous study for clean systems~\cite{Vagov2016, Wolf2017A, Cavalcanti2020, Vagov2020}, we can conclude that the IT vortex matter in dirty two-band superconductors with sufficiently large ratios of the band diffusion coefficients exhibits the formation of vortex clusters, and vortex chains in the IT subdomain above $\kappa^*_s(T)$ while vortex liquid droplets proliferate in the IT subdomain below $\kappa^*_s(T)$. According to the conclusions of \cite{Vagov2016, Wolf2017A, Cavalcanti2020, Vagov2020}, the appearance of such exotic vortex configurations is connected with the self-dual nature of the B-point~\cite{Vagov2016}. 

In summary, we have considered the nontrivial disorder effect appearing due to the interplay between the diffusive motion of charge carriers and the multiband structure of the single-particle states. Our results demonstrate that when the band diffusion coefficient in the weaker band is significantly larger than that of the stronger band, the nonmagnetic impurity scattering leads to a huge expansion of the IT domain between the standard superconductivity types in the $\kappa$-$T$ plane. In our study we have considered the minimal two-band diffusive model with the $s$-wave pairing in both bands that are coupled via Josephson-like interband pair transfer, while the interband impurity scattering is not included. However, our preliminary study makes it possible to conclude that the effect of interest is generic and the qualitative results are not sensitive to the interband scattering. Furthermore, the $s$-wave pairing is not crucial for our conclusions. Notice that the B-point is also present in the case of the $d$-wave pairing. Finally, our findings complement the previous results on the enlargement of the IT domain in clean multiband superconductors that takes place when the Fermi velocity of the weaker band is significantly larger than that of the stronger band~\cite{Vagov2016, Cavalcanti2020}.  

\subsection*{Acknowledgements}

The work was supported by the Basic Research Program of the HSE University.

\setcounter{equation}{0}
\renewcommand{\theequation}{S\arabic{equation}}
\renewcommand{\thefigure}{S\arabic{figure}}
\renewcommand{\bibnumfmt}[1]{[S#1]}
\renewcommand{\citenumfont}[1]{S#1}

\pagebreak
\twocolumngrid
\begin{center}
\textbf{\large Supplemental material for the article "Intertype superconductivity evoked by the interplay of disorder and multiple bands".}
\end{center}

\section{Expansion in powers of 
the band gap functions and their gradients}

\subsection{Expansion of the Green functions}

The first step in the derivation of the $\tau$-expansion ($\tau=1-T/T_c$, the proximity to the critical temperature) for the diffusive superconductor two-band model~\cite{Gurevich2003}, is the expansion of the Green functions $g_{\nu}({\bf r},\omega)$ and $f_{\nu}({\bf r},\omega)$ in powers of the band gap function $\Delta_{\nu}({\bf r})$ and its gradients. This expansion is sought in the form
\begin{align}
g_{\nu}=&g_{\nu}^{(0)} + g_{\nu}^{(2)} + g_{\nu}^{(4)} + \ldots,\notag\\
f_{\nu}=&f_{\nu}^{(1)} + f_{\nu}^{(3)} + f_{\nu}^{(5)} + \ldots,
\label{expgf}
\end{align}
where $g_{\nu}^{(2n)}$ and $f_{\nu}^{(2n+1)}$ are of the orders of $\Delta_{\nu}^{2n}$ and $\Delta_{\nu}^{2n+1}$, respectively (with $n=0,1,2,\ldots$). One also keeps in mind that the spatial gradient $\boldsymbol{\nabla}$ and gauge-invariant spatial derivative ${\bf D}=\boldsymbol{\nabla} -(i2e/\hbar \cbb){\bf A}$ are of the order of $\Delta_{\nu}$. Inserting Eq.~(\ref{expgf}) in the two-band Usadel equations [see Eq.~(1) in the article]  and using the accompanying normalization condition $g^2_{\nu} + |f_{\nu}|^2=1$, one obtains
\begin{align}
g_{\nu}^{(0)}=&1,\; g_{\nu}^{(2)}=-\frac{|\Delta_{\nu}|^2}{2(\hbar\omega)^2},\notag\\ g_{\nu}^{(4)}=&\frac{3|\Delta_{\nu}|^4}{8(\hbar\omega)^4}-\frac{{\cal D}_{\nu}}{2\hbar^2\omega^3}\mathrm{Re}\left(\Delta_{\nu}^*{\bf D}^2\Delta_{\nu}\right) \label{expg}
\end{align}
and 
\begin{align} 
f_{\nu}^{(1)}=&\frac{\Delta_{\nu}}{\hbar\omega_n},\; f_{\nu}^{(3)}=\frac{{\cal D}_{\nu}}{2\hbar\omega^2}{\bf D}^2\Delta_{\nu}-\frac{\Delta_{\nu}|\Delta_{\nu}|^2}{2(\hbar\omega)^3},\notag\\
f_{\nu}^{(5)}=&\frac{3\Delta_{\nu}|\Delta_{\nu}|^4}{8(\hbar\omega)^5}+\frac{{\cal D}^2_{\nu}}{4\hbar\omega^3} {\bf D}^2({\bf D}^2\Delta_{\nu})
\notag\\
&-\frac{{\cal D}_{\nu}}{4\hbar^3\omega^4}\Big[
3|\Delta_{\nu}|^2{\bf D}^2\Delta_{\nu}+
2\Delta_{\nu}|{\bf D}\Delta_{\nu}|^2 \notag\\
&+2\Delta_{\nu}^*({\bf D}\Delta_{\nu})^2+
\Delta_{\nu}^2({\bf D}^2\Delta_{\nu})^*\Big].
\label{expf}
\end{align}
We remark that only the terms contributing to the leading correction to the Ginzburg-Landau (GL) theory are highlighted in Eqs.~(\ref{expgf})-(\ref{expf}). 

\subsection{Expansion of the free energy density}

Here we outline the derivation of the free-energy expansion in powers of the band-dependent gap functions and their spatial gradients. To get this expansion, one inserts Eqs.~(\ref{expgf})-(\ref{expf}) in Eq.~(4) of the article. However, before proceeding to this calculation, we need to rearrange the last term in the brackets of Eq~(4). 

First, we obtain
\begin{align}
\int \frac{\hbar \cal{D}_{\nu}}{2}&\left[|{\bf D}f_{\nu}|^2+(\boldsymbol{\nabla} g_{\nu})^2\right] d^3{\bf r}\notag\\
&=\int \frac{\hbar \cal{D}_{\nu}}{2}\big[ - f^*_{\nu} {\bf D}^2f_{\nu} - g_{\nu} \boldsymbol{\nabla}^2 g_{\nu}\big] d^3{\bf r},
\label{step1}
\end{align}
where surface integrals, obtained by virtues of Gauss's theorem, vanish. Then, using the Usadel equation [see Eq.~(1) in the article], one gets   
\begin{align}
\int & \frac{\hbar \cal{D}_{\nu}}{2} \left[|{\bf D}f_{\nu}|^2+(\boldsymbol{\nabla} g_{\nu})^2\right] d^3{\bf r}\notag\\
&=\int \left[f^*_{\nu} \Delta_{\nu} + \hbar\omega g_{\nu}-\frac{2\hbar\omega+\hbar{\cal D}_{\nu} \boldsymbol{\nabla}^2 g_{\nu} }{2g_{\nu}} \right]d^3{\bf r}.
\label{step2}
\end{align}
In addition, to keep only the terms up to the order $\tau^3$ in the free energy, it is enough to use the approximations
\begin{align}
\int d^3{\bf r}\frac{\boldsymbol{\nabla}^2g_{\nu}}{g_{\nu}}=&\int d^3{\bf r}\frac{(\boldsymbol{\nabla} g_{\nu})^2}{g_{\nu}^2} \simeq\int d^3{\bf r} (\boldsymbol{\nabla} g_{\nu})^2,\notag\\
g_{\nu} \simeq &1-\frac{|f_{\nu}|^2}{2}-\frac{|f_{\nu}|^4}{8}-\frac{|f_{\nu}|^6}{16},
\label{step3}
\end{align}
where the latter expression follows from the normalization condition. Finally, we get the approximate expression
\begin{align}
\int \frac{\hbar \cal{D}_{\nu}}{2}&\left[|{\bf D}f_{\nu}|^2+(\boldsymbol{\nabla} g_{\nu})^2\right] d^3{\bf r}\notag\\
\simeq &\int d^3{\bf r} \bigg[f^*_{\nu} \Delta_{\nu}\notag + \hbar\omega\bigg(|f_{\nu}|^2 +\frac{|f_{\nu}|^4}{2}\notag\\ &+\frac{3|f_{\nu}|^6}{8}\bigg)-\frac{\hbar \cal{D}_{\nu}}{2} (\boldsymbol{\nabla} g_{\nu})^2\bigg],
\label{step4}
\end{align}

Now, inserting Eqs.~(\ref{expgf})-(\ref{expf}) and (\ref{step4}) in Eq.~(4) of the article, we get
\begin{align}
\fbb_{\nu}=&2\pi T N_{\nu}\sum\limits_{\omega >0}\bigg\{ -\frac{|\Delta_{\nu}|^2}{\hbar\omega} + \frac{|\Delta_{\nu}|^4}{4(\hbar\omega)^3}
-\frac{|\Delta_{\nu}|^6}{8(\hbar\omega)^5}\notag\\
&+ \frac{\hbar {\cal D}_{\nu}}{2(\hbar\omega)^2} |{\bf D}\Delta_{\nu}|^2
-\frac{(\hbar {\cal D}_{\nu})^2}{4(\hbar\omega)^3} |{\bf D}^2\Delta_{\nu}|^2 
\notag\\
&-\frac{\hbar {\cal D}_{\nu}}{8(\hbar\omega)^4}\big[6|\Delta_{\nu}|^2 |{\bf D}\Delta_{\nu}|^2+\Delta^2_{\nu}({\bf D}^*\Delta^*_{\nu})^2\notag\\
&+\Delta^{*2}_{\nu}({\bf D}\Delta_{\nu})^2\big]\bigg\}.
\label{fnu}
\end{align}
Then, we calculate the sums over $\omega$ as 
\begin{align}
&2\pi T\sum_{\omega>0}\frac{1}{\hbar\omega}=\ln\frac{2e^{\gamma}\hbar\omega_D}{\pi T},\;
2\pi T\sum_{\omega>0}\frac{1}{(\hbar\omega)^2}=\frac{\pi}{4T},\notag\\
&2\pi T\sum_{\omega>0}\frac{1}{(\hbar\omega)^3}=\frac{7\zeta(3)}{4\pi^2T^2},\; 2\pi T\sum_{\omega>0}\frac{1}{(\hbar\omega)^4}=\frac{\pi}{48 T^3},\notag\\
&2\pi T\sum_{\omega>0}\frac{1}{(\hbar\omega)^5}=\frac{31\zeta(5)}{16 \pi^4 T^4},
\label{sums}
\end{align}
where $\omega_D$ is the Debye frequency, $\zeta(\ldots)$ is the Riemann zeta function, and $\gamma=0.577$. 

Finally, using Eqs.~(\ref{fnu}) and (\ref{sums}), one gets Eq.~(5) of the manuscript. Notice that only the terms that contribute to the leading correction to the GL theory (in the $\tau$-expansion) are retained in this equation.  

\section{The $\tau$-expansion}

\subsection{The free energy functional}

Now we employ the $\tau$-expansion of the gap functions and fields in the form [$\vec{\Delta}^T=(\Delta_1,\Delta_2)$] 
\begin{align}
    &\vec{\Delta} = \tau^{1/2} \vec{\Psi}+ \tau^{3/2}\vec{\psi}+ \tau^{5/2}\vec{\varphi}+\ldots,\notag\\
    &{\bf A} = \tau^{1/2} \vecA + \tau^{3/2} \boldsymbol{\mathfrak{a}} + \dots,\notag\\ 
    &{\bf B} = \tau \vecB + \tau^2  \boldsymbol{\mathfrak{b}} + \dots,    
\label{tau}
\end{align} 
where $\vec{\Psi}=(\Psi_1,\Psi_2)^T$ and $\vecB$~($\vecA$) are the GL contributions to the gap functions and fields while the leading corrections are governed by $\vec{\psi}=(\psi_1,\psi_2)^T$ and $\mathfrak{b}$~($\mathfrak{a}$). Below we need also the next-to-leading correction to the gap functions $\vec{\varphi}=(\varphi_1,\varphi_2)$, which are not introduced in the article. The point is that $\vec{\varphi}$ appears in the leading correction to the GL theory in the $\tau$-expansion of the free energy functional but does not contribute to the stationary free energy. In addition, we introduce the $\tau$-scaling of the spatial coordinates; for more detail, see the discussion after Eq.~(7) in the article and see also the papers about the extended GL formalism for clean systems~\cite{Vagov2012A, Vagov2012B, Vagov2016, Cavalcanti2020, Vagov2020}. 

The corresponding $\tau$-expansion of the free energy density is written as
\begin{equation}
\fbb=\tau^2\left[\tau^{-1}\fbb^{(-1)} +\fbb^{(0)} + \tau \fbb^{(1)}+\ldots\right].
\label{ftau}
\end{equation}
For the lowest-order term we have
\begin{equation}
\fbb^{(-1)}=\vec{\Psi}^{\dag}\check{L}\vec{\Psi},
\label{f-1}
\end{equation}
where the matrix $\check{L}$ is given by Eq.~(8) of the article. The next order is given by 
\begin{equation}
\fbb^{(0)}=\frac{\vecB^2}{8\pi}+ \big(\vec{\Psi}^{\dag}\check{L}\vec{\psi} +\rm{c.c.}\big) +\sum\limits_{\nu=1,2} \fbb^{(0)}_{\nu}, 
\label{f0}
\end{equation}
where
\begin{equation}
\fbb^{(0)}_{\nu}= a_{\nu}|\Psi_{\nu}|^2+\frac{b_{\nu}}{2} |\Psi_{\nu}|^4 +{\cal K}_{\nu} |\vecD \Psi_{\nu}|^2,
\label{fnu0}
\end{equation}
with $\vecD=\boldsymbol{\nabla} - (i2e/\hbar \cbb)\vecA$. The coefficients $a_{\nu}, b_{\nu}$, and ${\cal K}_{\nu}$ are defined by Eq.~(6) of the article. 

Finally, $\frm^{(1)}$ is given by  
\begin{align}
\fbb^{(1)}=&\frac{\vecB \cdot \mathfrak{b}}{4\pi}+ \big(\vec{\Psi}^{\dag}\check{L}\vec{\varphi}+{\rm c.c.}\big) +\vec{\psi}^{\dag}\check{L}\vec{\psi}\notag\\
&+ \sum\limits_{\nu=1,2} \big[\fbb^{(1)}_{\nu,1}+ \fbb^{(1)}_{\nu,2}\big],
\label{f1}
\end{align}
where
\begin{align}
\fbb^{(1)}_{\nu,1}= & \frac{a_{\nu}}{2}\, |\Psi_{\nu}|^2 + b_{\nu}|\Psi_{\nu}|^4 + {\cal K}_{\nu}\,|\vecD \Psi_{\nu}|^2-{\cal Q}_{\nu} |\vecD^2 \Psi_{\nu}|^2\notag\\
&-\frac{{\cal L}_{\nu}}{2} \Big[6\,|\Psi_{\nu}|^2
|\vecD \Psi_{\nu}|^2 + \Psi_{\nu}^2
(\vecD^{\ast}\Psi_{\nu}^{\ast})^2\notag\\
&+\Psi_{\nu}^{\ast2}(\vecD\Psi_{\nu})^2\Big] - \frac{c_{\nu}}{3}|\Psi_{\nu}|^6
\label{f1nu1}
\end{align}
and 
\begin{align}
\fbb^{(1)}_{\nu,2}=&\big(a_{\nu}+ b_{\nu} |\Psi_{\nu}|^2\big)\big(\Psi_{\nu}\psi_{\nu}^*+{\rm c.c.}\big)\notag\\ 
&+{\cal K}_{\nu}\left[\big(\vecD \Psi_{\nu}\cdot \vecD^*\psi^*_{\nu}
+ {\rm c.c.}\big) - \mathfrak{a} \cdot \boldsymbol{\mathfrak{i}}_{\nu} \right], 
\label{f1nu2}
\end{align}
where $\boldsymbol{\mathfrak{i}}_{\nu} = 4e {\rm Im} [\Psi_{\nu}^* \vecD\Psi_{\nu}]/{\hbar \cbb}$.

Using the $\tau$-expansion of the free energy functional given by Eqs.~(\ref{ftau})-(\ref{f1nu2}), one gets the stationary point equations, see Eqs.~(5) and (6) in the article. According to Eq.~(5), we find that the contribution $\fbb^{(-1)}$ is exactly equal to zero at the stationary point. In addition, the term $\vec{\Psi}^{\dag}\check{L}\vec{\varphi}+{\rm c.c.}$ is also zero in $\frm^{(1)}$. Thus, only $\Psi$ and $\psi$ make a contribution to the free energy density up to the order $\tau^3$. Moreover, $\psi$ is written as the linear combination of $\vec{\xi}$ and $\vec{\eta}$ as $\vec{\psi}=\psi_{\xi}\vec{\xi} + \psi_{\eta} \vec{\eta}$, see also Eq.~(13) in the article. Then, one can find~\cite{Vagov2012B, Vagov2016, Cavalcanti2020} that only $\psi_{\eta}$ contributes to the leading correction to the GL theory and furthermore, $\psi_{\eta}$ is expressed in terms of $\Psi$, see Eqs.~(15) and (16) in the main article. Thus, to find the stationary free energy up to the leading correction to the GL theory (this corrections is of the order of $\tau^3$), one needs to use only the stationary solution of the GL formalism. 

\subsection{The Gibbs free energy difference}

Using the stationary free energy, one calculates the Gibbs free energy difference given by Eq.~(18) in the article. To simplify the calculations, we introduce the dimensionless quantities
\begin{align}
    &\tilde {\bf r} = \frac{\bf r}{\lambda \sqrt{2}},\,\tilde \vecB = \frac{\kappa\sqrt{2}}{\mathcal{H}_c} \vecB,\, 
    \tilde \vecA = \frac{\kappa}{\mathcal{H}_c \lambda} \vecA,\notag\\
    &\tilde \Psi = \frac{\Psi}{\Psi_0},\; 
    \tilde{\fbb} = \frac{4\pi\,\fbb}{\mathcal{H}_c},\; 
    \tilde{\gbb} = \frac{4\pi\,\gbb}{\mathcal{H}_c},
\label{dim_var}
\end{align}
where $\Psi_0=\sqrt{-a/b}$ is the uniform solution of the GL formalism and $\mathcal{H}_c$ is the GL thermodynamic critical field, see Eq.~(19) in the article. Below we utilize these dimensionless quantities without tilde, for simplicity. Notice that we use the $\tau$-scaled spatial coordinates and so, the GL coherence length $\xi$ and the London penetration depth $\lambda$ are scaled accordingly.

The series in $\tau$ for the Gibbs free energy difference is sought in the form 
\begin{align}
\gbb = \tau^2\big[\gbb^{(0)} + \tau \gbb^{(1)} +\dots\big], 
\label{gtau}
\end{align}
where the lowest order (GL) contribution is given by 
\begin{align}
\gbb^{(0)}=\frac{1}{2}\left(\frac{\mathcal{B}}{\kappa\sqrt{2}}
- 1 \right)^2+\frac{1}{2\kappa^2}\,|\vecD\Psi|^2 - |\Psi|^2 + \frac{1}{2}\,|\Psi|^4, \label{g0}
\end{align}
where $\mathcal{B}=|\vecB|$ and the dimensionless gauge-invariant derivative is given by $\vecD=\boldsymbol{\nabla} +i\vecA$. The leading correction to the GL theory reads
\begin{align}
\gbb^{(1)}=&\left(\frac{\mathcal{B}}{\kappa\sqrt{2}}-1\right) \left[\frac{1}{2}+ \bar{c} + \bar{\mathcal{G}}(\bar{\alpha} -\bar{\beta})^2\right] + \bar{\mathcal{G}}|\Psi|^2\notag\\ &\times \big(\bar{\alpha}-\bar{\beta}|\Psi|^2\big)^2- \frac{1}{2}|\Psi|^2+|\Psi|^4+\frac{1}{2\kappa^2}|\vecD \Psi|^2\notag\\
&+ \frac{\bar{\cal Q}}{4\kappa^4}|\vecD^2\Psi|^2+\frac{\bar{\cal L}}{4\kappa^2} \Big\{6|\Psi|^2 |\vecD \Psi|^2 \notag\\
&+\big[\Psi^2(\vecD^*\Psi^*)^2 + {\rm c.c.}\big]\Big\} +\bar{c}\,\big|\Psi\big|^6, 
\label{g1}
\end{align}
where the dimensionless coefficients $\bar{c}, \bar{Q}, \bar{L}$, and $\bar{\mathcal{G}}$ are given by Eq.~(21) of the article. 

Using Eqs.~(\ref{gtau})-(\ref{g1}) and introducing the expansion in $\delta\kappa=\kappa-\kappa_0$~(with $\kappa=\kappa_0=1/\sqrt{2}$, see the article), one gets Eq.~(20). This makes it possible to employ the self-duality Bogomolnyi equations~\cite{Vagov2016, Cavalcanti2020}, as the the GL theory is reduced to these equations at $\kappa=\kappa_0$. The relevant details of the perturbation expansion in $\delta\kappa$ and the corresponding calculations are discussed in the previous papers on the IT domain in clean superconductors~\cite{Vagov2016, Cavalcanti2020}.

\end{document}